\documentclass[usenatbib]{mnras}

\usepackage{newtxtext,newtxmath}

\usepackage[T1]{fontenc}
\usepackage{ae,aecompl}

\usepackage{graphicx}	
\usepackage{amsmath}	
\usepackage{amssymb}	

\title[Optically thin outbursts of neutron stars]{Optically thin outbursts of rotating neutron stars can not be spherical}

\author[M. Wielgus]{
Maciek Wielgus$^{1,2}$\thanks{E-mail: mwielgus@cfa.harvard.edu}
\\
$^{1}$Black Hole Initiative at Harvard University, 20 Garden Street, Cambridge, MA 02138, USA \\
$^{2}$Harvard-Smithsonian Center for Astrophysics, 60 Garden Street, Cambridge, MA 02138, USA
}

\date{Accepted XXX. Received YYY; in original form ZZZ}

\pubyear{2019}

\begin{document}
\label{firstpage}
\pagerange{\pageref{firstpage}--\pageref{lastpage}}
\maketitle

\begin{abstract}
We investigate three-dimensional relativistic trajectories of test particles in the spacetime of a~slowly rotating compact star, under the combined influence of gravity and a~strong, near-Eddington radiation field. While in the static case a~spherically symmetric shell of matter suspended above the stellar surface can be formed at the location of radial equilibrium of effective forces, the same is not true for a~rotating star. In the latter case the symmetry is broken by the interplay between motion in the~non-static spacetime and the influence of strong radiation drag forces, pushing particles towards the equatorial plane. As a~result an expanding spherical shell of matter ejected from the neutron star surface collapses on a short timescale into a~single stable equatorial ring supported by radiation. These findings have implications for the geometry of optically thin outflows during luminous neutron star bursts.
\end{abstract}




\section{Introduction}

During type-I X-ray burst neutron stars may
reach near-Eddington luminosities  \citep{Tawara1984,Lewin1993}, leading to photospheric radius expansion events \citep{Lewin1984, Strohmayer2006}. This type of outburst is of significant importance for studying neutron stars, including the estimation of their equation of state \citep{Ozel2006,Lattimer2012}, spin \citep{Muno2001}, mass and radius \citep{Galloway2008, Bollimpalli2019}, and distance \citep{Kuulkers2003, Galloway2003}. It is common to assume spherical symmetry of such events \citep{Bildsten2000, Strohmayer2019}, which is an important constraint for the considered models of expansion. Most of the models ignore the dynamical effects of the strong radiation field such as Poynting-Robertson drag \citep[however, see][]{Worpel2013}.

To understand the fundamental aspects of the dynamics of matter during a~near-Eddington luminosity outburst on a~neutron star, \citet{Abramowicz1990} and \citet{Miller1996} initiated studies of the influence of a~strong radiation field on general relativistic trajectories of particles. More recently this problem has received a~lot of attention from the community, with multiple groups addressing essential questions related to the location of the radiation-gravity equilibrium surface \citep{SokOh2011, Stahl2012}, a~relativistic description of the radiative drag forces \citep{Bini2009, Mishra2014} and motion under different configurations of the emitting surface \citep{Bini2015,Wielgus2015}, including the influence of rotation \citep{Bini2011}. While most works assume spherical symmetry \citep{Wielgus2012, Stahl2013} or confine the analysis to the equatorial plane of an axisymmetric spacetime \citep{SokOh2010}, three dimensional trajectories have been studied recently as well \citep{DeFalco2019}. In parallel, efforts were made to extend this work from studying test particles' trajectories to a~description of equilibrium states of optically thin \citep{Wielgus2015b} and optically thick \citep{Wielgus2016} gas, and dynamics of gaseous envelopes \citep{Bollimpalli2017}. The relativistic dynamical effects of radiation are difficult to trace with global (magneto)hydrodynamical simulations, which have significant difficulties reproducing the exact form of the radiation field \citep[see, e.g.,][]{Sadowski2013}. However, attempts are being made \citep{Fragile2018}.

In this work we consider three-dimensional relativistic trajectories of particles in the vicinity of a~slowly rotating luminous neutron star, consistently treating the geometry of spacetime and of the radiation source. We show that the particles are strongly affected by radiation drag, forcing them into the neutron star's equatorial plane on a~relatively short timescale. This effect implies that no optically thin envelope supported by radiation can form around a~rotating neutron star.

\section{Equations of motion}

The relativistic equation of motion for a~test particle of mass $m$ accelerating under influence of the radiation four-force $F^\mu$ can be written as \citep[see, e.g.,][]{Stahl2013}
\begin{equation}
\label{eq:eom}
a^\mu = u^\nu \nabla_\nu u^\mu = \frac{1}{c^2}\frac{\rm d^2 x^\mu}{\rm d \, \tau^2 \,} + \Gamma^\mu_{\ \nu \rho}u^\nu u^\rho = \frac{\sigma_{\rm T}}{m c^3} F^\mu \ .
\end{equation}
Hereafter $a^\mu$ represents the particle's four-acceleration, $u^\mu$ its four-velocity, $x^\mu$ is the position of the particle in a~chosen system of coordinates with associated Christoffel symbols of the second kind $\Gamma^\mu_{\ \nu \rho}$ and $\tau$ is the proper time. For simplicity we assume a~pure hydrogen plasma, that is, $m$ is a~proton mass and $\sigma_{\rm T}$ is the~Thomson cross section. Using the projection tensor $h^\mu_{\ \nu}$, Kronecker tensor $ \delta^\mu_{ \ \nu}$, and the radiation stress-energy tensor $T^{\nu \rho}$, the radiation four-force can be represented as
\begin{equation}
\label{eq:force}
    F^\mu = h^\mu_{\  \nu} T^{\nu \rho}u_\rho \equiv -\left( \delta^\mu_{ \ \nu} + u^\mu u_\nu \right) T^{\nu \rho}u_\rho \ .
\end{equation}
A detailed description of motion in an astrophysically relevant situation requires a~realistic model of the radiation field. An analytic radiation stress-energy tensor $T^{\hat{\mu} \hat{\nu}}$ for a~static, uniformly radiating luminous neutron star of mass $M$, radius $R$ and luminosity seen by a~distant observer $L_\infty = \lambda L_{\rm{Edd}}$ was first given by \cite{Abramowicz1990}. The dimensionless parameter $\lambda$ represents the luminosity in Eddington units, where $ L_{\rm{Edd}} = 4\pi m GM c / \sigma_{\rm T} $.
\cite{Miller1996} proved that for a~slowly rotating star, with dimensionless angular momentum $|j| \ll 1$, the local non rotating frame (LNRF) components found for the static case remain approximately valid, and new non-zero components, $T^{\hat{t} \hat{\phi}}$ and $T^{\hat{r} \hat{\phi}}$, appear. The full radiation stress-energy tensor, valid to first order in $j$, takes the following form in $G = c = 1$ units
\begin{align}
    \label{eq:Ttt}
    T^{\hat{t} \hat{t}} &= 2 \pi I \left( 1 - \cos \alpha_0 \right) \ , \\
    T^{\hat{t} \hat{r}} &= \pi I \sin^2 \alpha_0 \ , \\
    T^{\hat{r} \hat{r}} &=  \frac{2 \pi}{3} I \left(1 - \cos^3 \alpha_0 \right) \ , \\
    T^{\hat{\phi} \hat{\phi}} &= T^{\hat{\theta} \hat{\theta}} =  \frac{\pi }{3} I \left(\cos^3 \alpha_0 - 3 \cos \alpha_0 + 2\right) \ , \\
    T^{\hat{t} \hat{\phi}} &= \frac{\pi }{ 3}  I J \left( \cos^3 \alpha_0 - 3 \cos \alpha_0 + 2 \right) \ ,\\
    T^{\hat{r} \hat{\phi}} &= \frac{\pi}{4} I J \sin^4 \alpha_0 \, ,
    \label{eq:Trf}
\end{align}
with uniform frequency-integrated specific intensity $I$,
\begin{equation}
I(r) = I(R) \left(\frac{1-\frac{2M}{R}}{1-\frac{2M}{r}} \right)^2 = \frac{m M (1-\frac{2M}{R})}{\pi \sigma_{\rm T} R^2 (1-\frac{2M}{r})^2} \lambda \, , 
\end{equation}
apparent viewing angle of a~static star $\alpha_0$,
\begin{equation}
\sin \alpha_0 =   \frac{R}{r} \left(\frac{1-2M/r}{1-2M/R} \right)^{1/2}  \ ,
\end{equation}
and special function $J(r)$,
\begin{align}
J(r) = & 8j \frac{r}{M} \left[\left(\frac{M}{R}\right)^3 - \left(\frac{M}{r}\right)^3 \right]\left(1 - \frac{2M}{r} \right)^{-1/2} \nonumber \\
&+ \frac{4}{\sin \alpha_0} \frac{j}{\pi}\frac{M}{R}\left(5 - 4\frac{M}{R} \right) \, .
\label{eq:Jr}
\end{align}
\begin{figure}
\centering
\includegraphics[width=0.999\columnwidth]{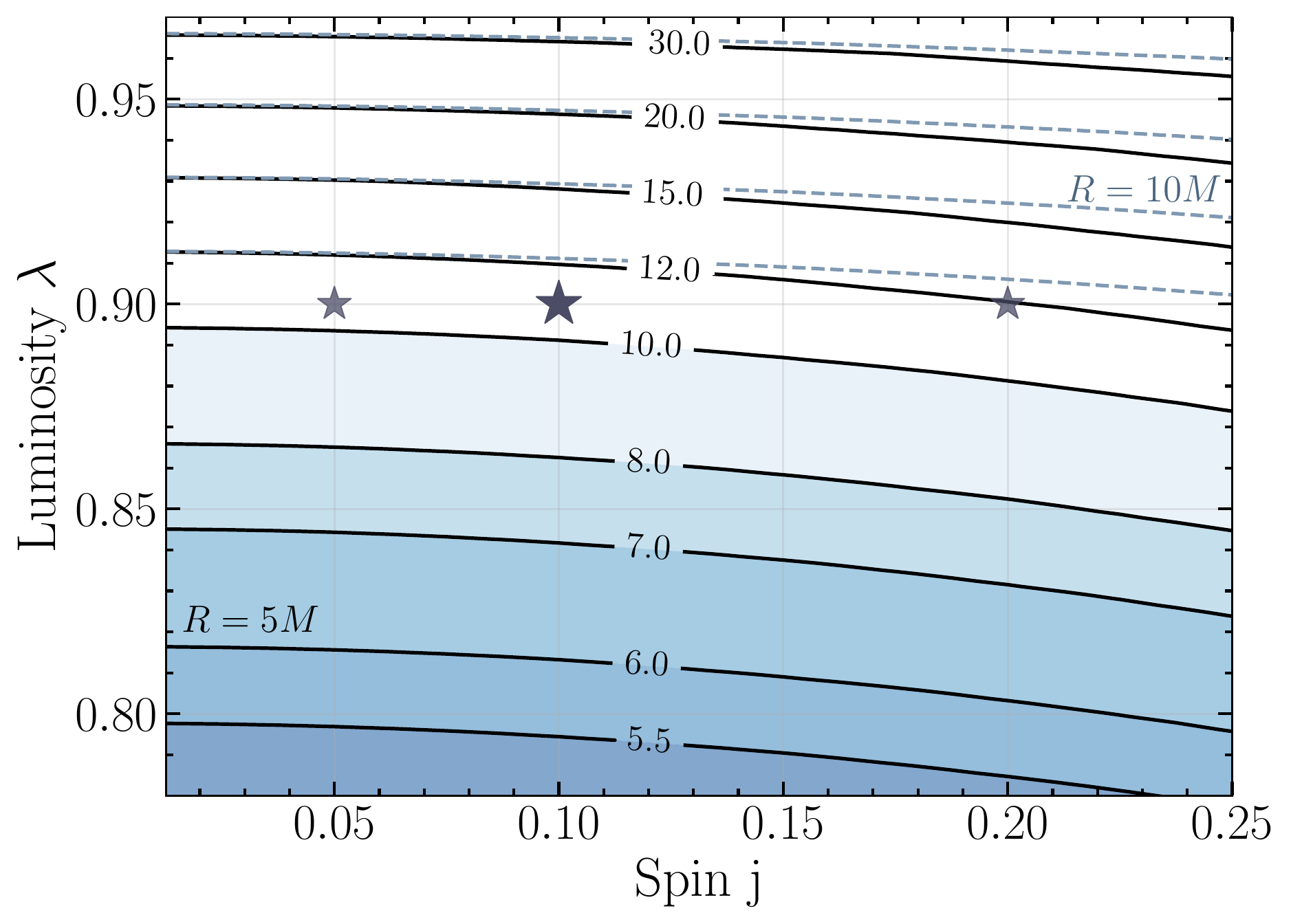}
\caption{
Contour plot of the stable circular orbit (Eddington capture ring, ECR) radius dependence on the parameters of luminosity $\lambda$ and spin $j$. The stellar radius $R = 5M$ is plotted with continuous lines, slightly deviating curves for $R=10M$ are plotted with dashed blue lines. For $j=0$ ECS radius, as described analytically by eq. (\ref{eq:ecs}), is recovered. Stars denote the specific cases presented in this paper.}
\label{fig:orbit}
\end{figure}
\begin{figure*}
\centering
\includegraphics[width=0.38\textwidth]{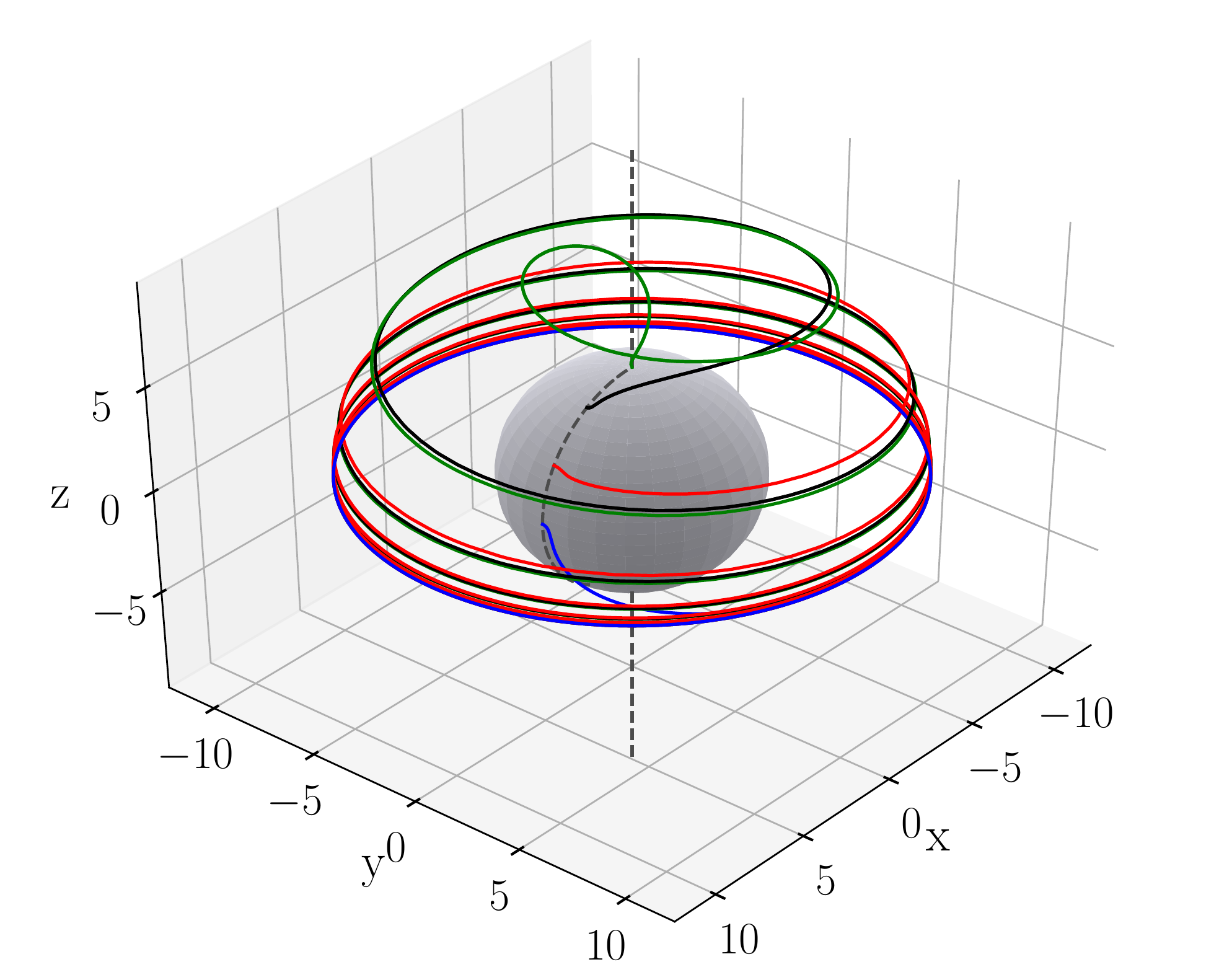}
\includegraphics[width=0.61\textwidth]{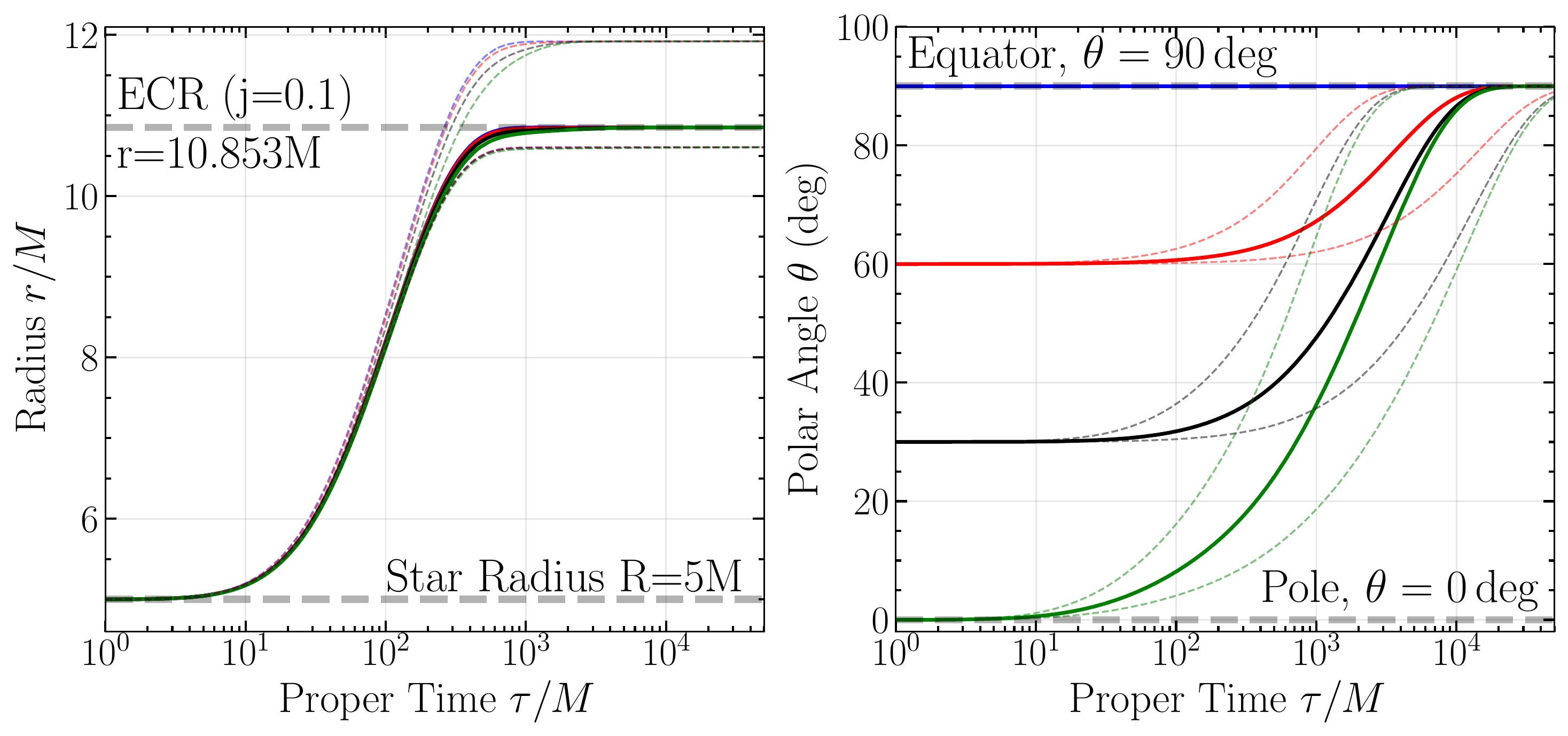}
\caption{ Left: trajectories of particles ejected from a~bursting neutron star with luminosity $\lambda=0.9$, spin $j=0.1$, and stellar radius $R=5M$, initially located at polar angles $0.0001, 30, 60$ and $90 \deg$. Middle: radial positions of particles shown in the left panel as functions of the proper time $\tau$. Similar trajectories for spin $j=0.2$ (larger critical radius) and $j=0.05$ (smaller critical radius) shown with thin dashed lines. Particles reach the critical sphere in $\tau \sim 10^3 M$. Right: polar positions of particles shown in the left panel as function of the proper time $\tau$. Similar trajectories for spin $j=0.2$ (faster equatorial drift) and $j=0.05$ (slower equatorial drift) shown with thin dashed lines. Particles reach the equatorial plane in $\tau \sim 10^4 M$.}
\label{fig:3d_trajectories}
\end{figure*}
The two components of the eq. (\ref{eq:Jr}) correspond to general relativistic frame-dragging, and Doppler frequency shift \citep{Miller1996}.
The approximation assumes that the emitting surface (and the whole star) rotates slowly and rigidly. This description allows us to consistently account for the angular momentum of photons emitted by an extended rotating source, in contrast to previous approaches of \cite{Bini2011}, assuming constant angular momentum or \cite{DeFalco2019}, assuming no angular momentum in the LNRF.
Since the calculations are only valid in the low spin approximation, we follow \citet{SokOh2010} and consider the Kerr spacetime simplified by only including terms linear in spin $j$, 
\begin{align}
    {\rm d} \tau^2  =& -(1-2M/r) {\rm d} t^2 + \frac{{\rm d}r^2}{1-2M/r} + r^2 {\rm d} \theta^2  \nonumber \\
    &+ r^2 \sin^2 \theta \left({\rm d} \phi^2 - 4 j \frac{M^2}{r^3} {\rm d}\phi {\rm d}t \right) \ ,
\label{eq:metric}
\end{align}
which is identical to the~low spin, zero quadrupole moment Hartle-Thorne metric \citep{Hartle1968}. To obtain the final equation of motion, we transform the LNRF tetrad components $T^{\hat{\mu}\hat{\nu}}$ to Boyer-Lindquist coordinates. The exact components of the transformation tensor $e^\mu_{\ \hat{\mu}}$ follow from \citet{Bardeen1970,Miller1996}. Finally, $T^{\mu \nu}$ can be plugged into the radiative force eq. (\ref{eq:force}) and then into the equation of motion (\ref{eq:eom}). We integrate eq. (\ref{eq:eom}) numerically, utilizing the four-velocity normalization $u^\mu u_\mu = -1$ as an additional constraint, to obtain three-dimensional trajectories of test particles.

\section{Results}

This work builds upon numerous previous findings:
\begin{enumerate}
\item For a non-rotating, non-luminous star we reproduce the Schwarzschild spacetime time-like geodesics \citep[e.g.,][]{Hartle2003},

\item For a rotating, non-luminous star we produce a variety of three-dimensional geodesic trajectories characteristic to axially symmetric spacetimes \citep[e.g.,][]{Grossman2012},

\item for a non-rotating luminous star we reproduce the results of \citet{SokOh2011, Stahl2012} and the presence of radial equilibrium, referred to as the Eddington capture sphere (ECS),

\item for equatorial motion around a~rotating luminous star we find that the trajectories of particles converge to a~single ring, which is consistent with the findings of \citep{SokOh2010}. The location of this ring is a~function of luminosity $\lambda$, stellar radius $R$ and spin $j$, however, the dependence on $R$ and $j$ is rather weak, see fig. \ref{fig:orbit},

\item in three dimensions, in the point source limit of $R \rightarrow 0$, for near-Eddington luminosity, we reproduce the presence of the critical surface reported by \cite{DeFalco2019}. Agreement in the slow drift of particles towards the equatorial plane required implementing the full Kerr metric (that is, including terms that are higher order in spin $j$ in the metric,  absent in eq. \ref{eq:metric}), which we have done as a~validation test. 
\end{enumerate}
For a~static source with near-Eddington luminosity there is a~sphere, located at the radius 
\vspace{2.5mm}
\begin{equation}
r_{\rm{ECS}} = \frac{2M}{1-\lambda^2} \ ,
\label{eq:ecs}
\end{equation}
for which the effective gravity and the radial component of the radiation force balance one another. Motions tangential to the ECS are damped by Poynting-Robertson drag, efficiently removing the particles' angular momentum and resulting in a~stable equilibrium surface \citep{Stahl2012}. As particles accumulate at the ECS, an optically thin levitating atmosphere can be formed. Because the effective force acting on optically thin piece of fluid is always directed towards the ECS, which is collocated with density and pressure maximum of the atmosphere, such configuration is Rayleigh-Taylor stable \citep{Wielgus2016}. This conclusion does not not generalize to optically thick atmospheres easily. As long as we are limited to the equatorial plane, including stellar rotation does not have any dramatic influence on the trajectories of the captured test particles -- instead of stabilizing at $r = r_{\rm{ECS}}$, particles slowly orbit the star on a~circular trajectory of radius $r \gtrapprox r_{\rm{ECS}}$, fig. \ref{fig:orbit}. Hereafter we refer to the locus of these equatorial trajectories as the Eddington capture ring (ECR). The constant orbital velocity in the coordinate frame comes from a~combination of the Doppler effect from the rotating extended stellar surface and the Lense-Thirring frame-dragging effect \citep{SokOh2010}. The picture looks dramatically different in three dimensions. Off-equatorial time-like geodesics in a~non-static axisymmetric spacetime cannot be confined to a~single plane and exhibit vertical precession around the equatorial plane \citep{Bardeen1972}. However, based on Schwarzschild spacetime intuition one may imagine near-Eddington luminosity to produce the radial force necessary to keep the particles on a~surface of a~spherical topology encompassing the neutron star, and radiation drag forces to stabilize motions in directions tangential to such a~surface. This notion is incorrect, at least in part. While a~surface of effective radial forces balance does exist, particles cannot remain static there because of the unbalanced effective tangential forces. Interplay between frame-dragging and the radiation force created by an extended rotating source results in tangential motions being damped, but stabilized uniquely in the equatorial plane $\theta = 90 \deg$, where $u^\theta=0$ implies $F^\theta = 0$ and $\rm{d}^2\theta /\rm{d} s^2 = 0$. Hence, as proper time $\tau$ grows, the motion is being reduced to the planar equatorial one discussed earlier.

\begin{figure}
\centering
\includegraphics[width=0.999\columnwidth]{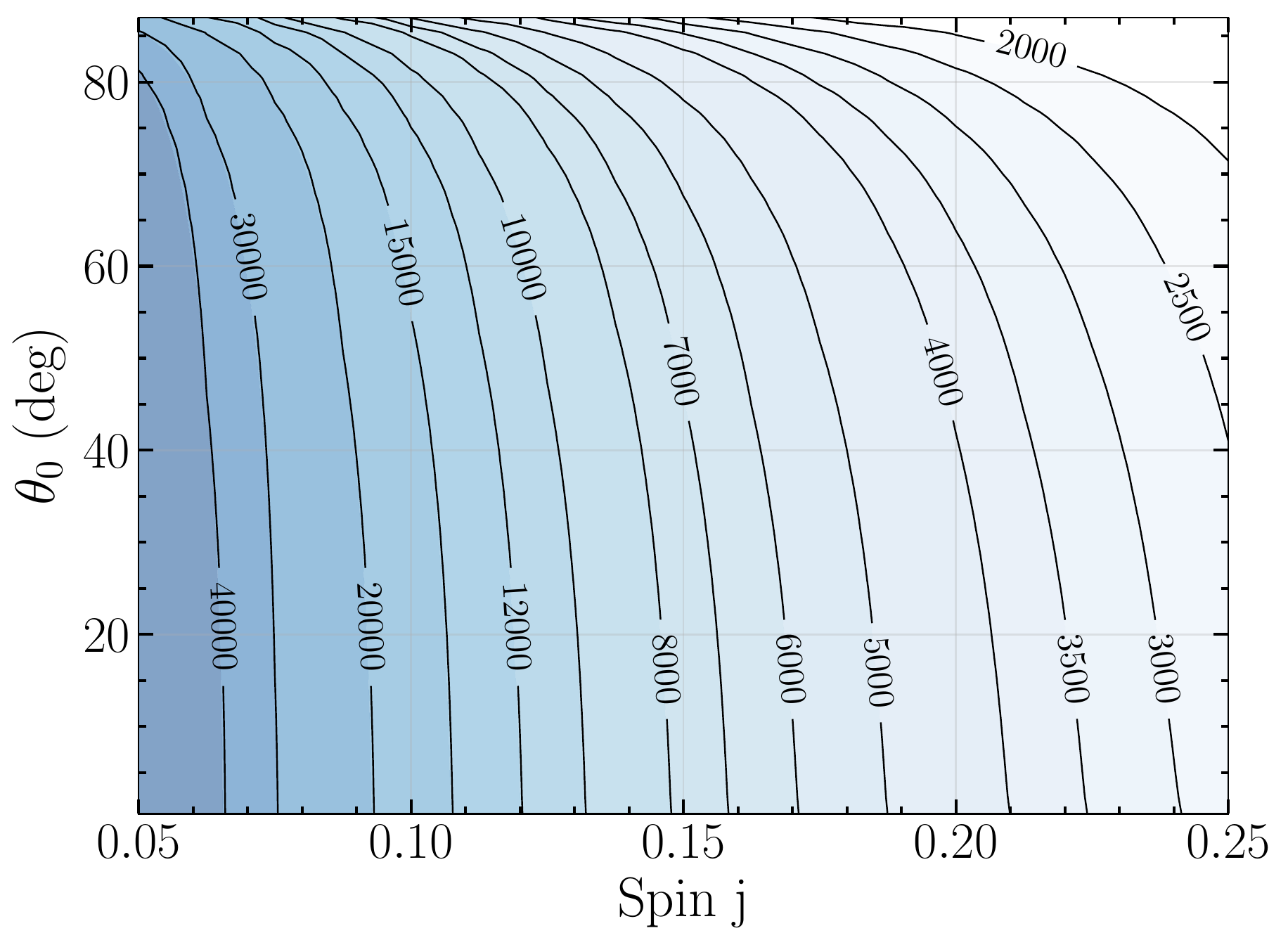}
\caption{ Contour plot of time $\tau_{\rm{set}}$ required for a particle ejected from the neutron star surface to settle down in the equatorial plane ECR, as a~function of the spin $j$ and particle's initial polar position on the neutron star surface $\theta_0$. Luminosity $\lambda=0.9$, star radius $R=5M$. Equatorial plane is located at $\theta = 90 \deg$.}
\label{fig:settle_time}
\end{figure}

\begin{figure*}
\centering
\includegraphics[width=0.38\textwidth]{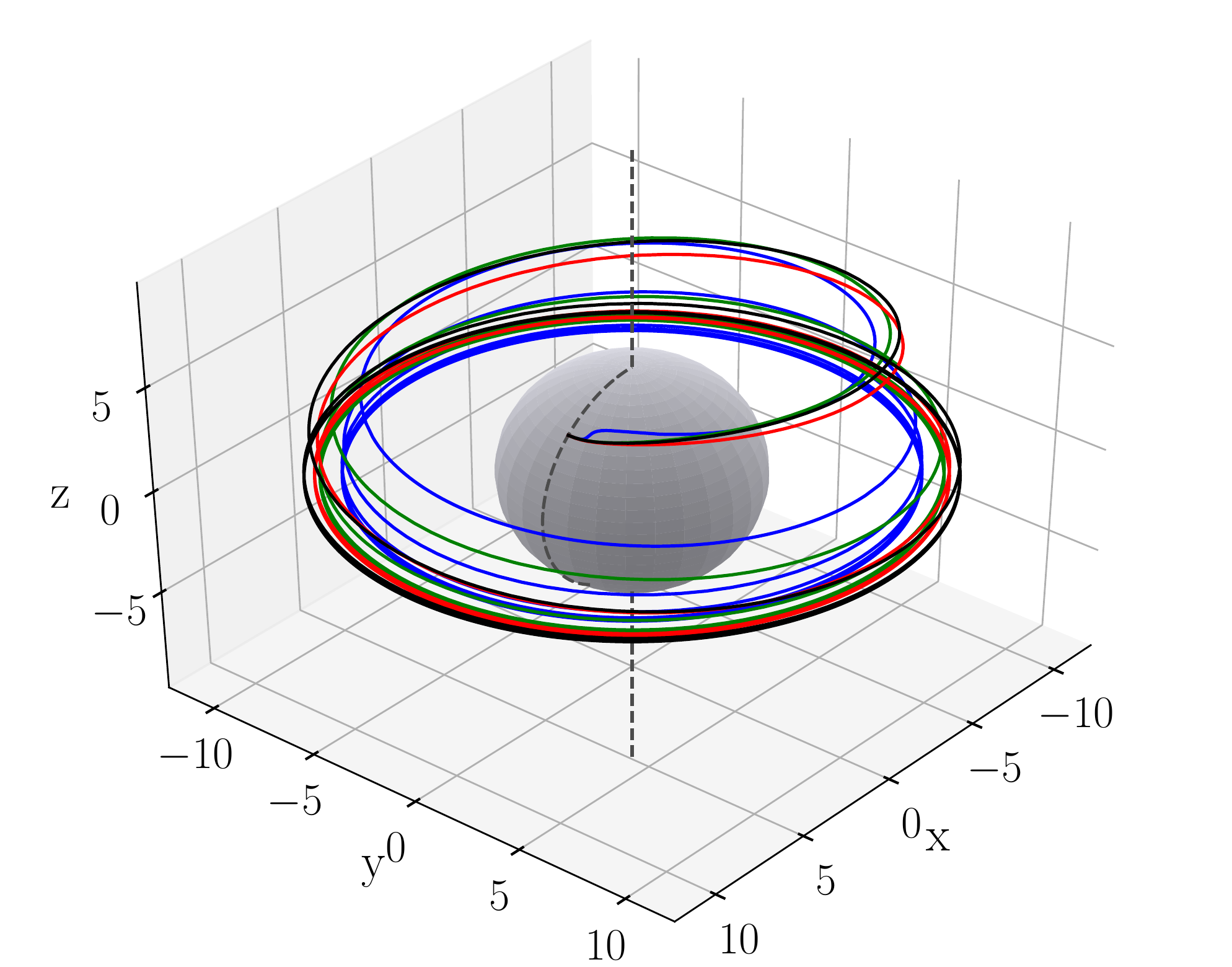}
\includegraphics[width=0.61\textwidth]{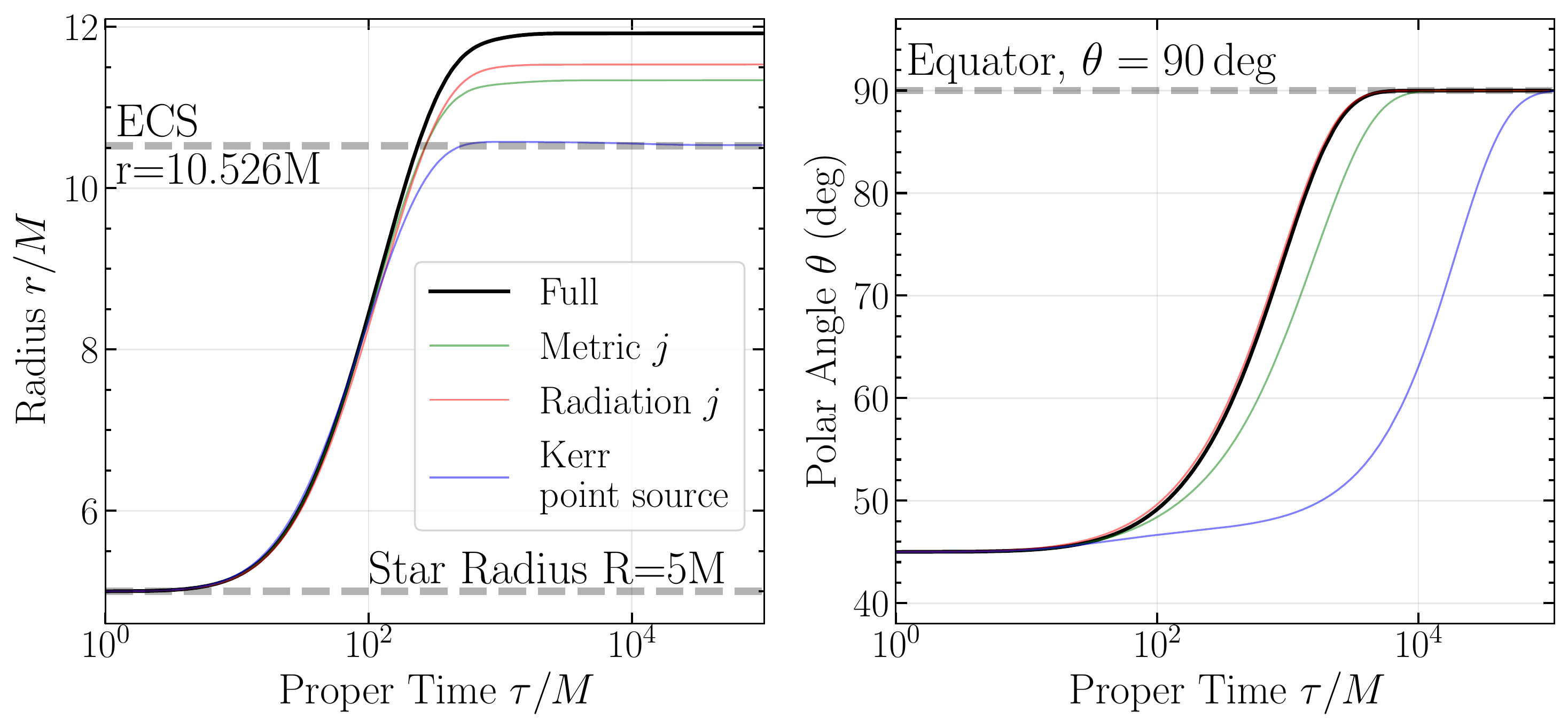}
\caption{ Left: trajectories of particles ejected from a~bursting neutron star with luminosity $\lambda=0.9$, spin $j=0.2$, and stellar radius $R=5M$, initially located at polar angle $\theta_0 = 45 \deg$ and corotating with the stellar surface, for different code variants: full code (black), spin included only in the metric (green), spin included only in the radiation tensor (red), point source radiation in Kerr metric (blue). Middle: radial positions of particles shown in the left panel as function of the proper time $\tau$ for the four code variants. Right: polar positions of particles shown in the left panel as a function of the proper time $\tau$ for the four code variants.}
\label{fig:3d_trajectories_comp}
\end{figure*}

\subsection{Burst from the neutron star surface}
\label{subs:burst_surface}
To illustrate the behavior described above, we consider an outburst of matter from the surface of a~neutron star, driven by a~rapid near-Eddington radiation burst. We consider a~star with radius $R=5M$ and luminosity instantaneously increased from $0$ to $\lambda=0.9$ at proper time $\tau_0=0$. Particles are initialized at the stellar surface with zero velocity in the corotating frame. Examples of trajectories are given in fig. \ref{fig:3d_trajectories}. In the left panel a~star rotating counter-clockwise is shown along with three-dimensional trajectories found for initial polar angles $\theta_0$ of $0.0001 \deg$ (near the rotation axis), $30$, $60$ and $90 \deg$. The middle and right panel of fig. \ref{fig:3d_trajectories} show radial and polar coordinates of the particles along their trajectories. Particles initially follow the rotating star surface, but pushed by the strong radial radiation flux, they quickly gain radial velocity. As the particles approach the critical surface and gravity starts to balance the radiation, the radial velocity drops down. Subsequently, the particles are forced to spiral towards the equatorial plane, to remain there on the stable ECR. Both the radius of the ECR and the time required for the particle to settle down in the equatorial plane depend on the neutron star spin $j$. 

We explore this equatorial settle-down time in fig. \ref{fig:settle_time}. Trajectories of particles for a~grid of spins $j$ and initial polar angles $\theta_0$ were computed, evaluating the time required for the particle to get closer than 0.01 radians to the equatorial plane in the polar direction. The characteristic settling time is $\tau_{\rm{set}} < 2 \times 10^4 M$ for spin $j= 0.1$ and $\tau_{\rm{set}} < 5 \times 10^3 M$ for spin $j= 0.2$. For a~neutron star of mass of $1.5 M_\odot$ we find $G M / c^3 \approx 7.5 \mu$s, yielding a settle-down time of $0.15$\,s and $0.04$\,s, respectively. While we have ignored special and relativistic corrections required to translate this time to the one measured by a~distant observer, it is clear that the equatorial settle-down time is significantly shorter than a~typical photospheric expansion burst duration, which is few seconds \citep{Strohmayer2006}.

\subsection{Non-static spacetime or source rotation?}
Is the presented effect related rather to the rotation of the luminous extended surface or to the rotation of the central mass? We attempt to answer this question by comparing trajectories found for three different versions of the integrated equations, under the same initial conditions, see fig. \ref{fig:3d_trajectories_comp}.
Trajectories found with the full numerical code are shown with thick black lines. Curves obtained for the case in which the spin term $j$ was exclusively included in the LNRF radiation tensor description in eq. (\ref{eq:Jr}) are shown with thin red lines. Finally, the version of equation of motion, for which spin term was included exclusively in the metric-related terms is represented with thin green lines. Both incomplete treatments recover qualitatively similar trajectories to the general case, with the final radii smaller by a~few percent. The prescription ignoring spin in the metric recovers the polar component of trajectories which are very similar to the full code (red and black curves are overlaid in the right panel of fig. \ref{fig:3d_trajectories_comp}), ignoring spin in the radiation tensor results in a~lagged equatorial drift. Regardless of whether the spherical symmetry is broken by the metric or by the radiation tensor field, particles are efficiently pushed towards the equator. However, a proper quantitative description of the particle's trajectory requires both components.

\subsection{Importance of the star finite size}
\label{sub:finite_size}
Finally, to quantify the importance of the non-zero source size, we consider test particle in Kerr metric, influenced by the radiation consistent with a point source, mimicking the radiation field description presented in \citet{DeFalco2019}. We find the relevant stress-energy tensor components by taking the stellar radius $R \rightarrow 0$ limit in eqs. (\ref{eq:Ttt})-(\ref{eq:Trf}). The only remaining non-zero components of the $T^{\hat{\mu}\hat{\nu}}$ tensor are
\begin{equation}
T^{\hat{t}\hat{t}} = T^{\hat{r}\hat{r}} = T^{\hat{t}\hat{r}} = \frac{mM \lambda}{ \sigma_{\rm{T}}r^2 (1-2M/r)} \, .
\label{eq:point_source_radiation}
\end{equation}
 Figure \ref{fig:3d_trajectories_comp} shows comparisons between trajectories obtained with the radiation tensor described by eq. (\ref{eq:point_source_radiation}) in Kerr spacetime (thin blue lines) and by eqs. (\ref{eq:Ttt})-(\ref{eq:Trf}) in the metric given by eq. (\ref{eq:metric}), black lines. It can be seen that point source radiation test resulted in a critical surface radius value close to the Schwarzschild metric ECS. This is because the radial balance equation only has a~second order dependence on spin in the $R \rightarrow 0$ limit. While the equatorial drift is still present even for the point source, it happens on a~timescale slower by an order of magnitude. This large discrepancy stresses how phenomena related to the finite source size, such as the rotating surface Doppler shift, are important to capture the relativistic dynamics of particles near the luminous stellar surface.

\section{Discussion}
The described effect will necessarily change geometry of an optically thin flow. As a~consequence, outflows driven by luminous (near-Eddington) bursts from compact, rotating objects can not maintain spherical geometry. Instead, the ejected matter is redistributed and forced to settle-down at the equatorial plane. Importantly, the timescale for this equatorial drift is significantly shorter than the characteristic duration of luminous bursts. While in this work we discuss test particles (dust), the results should straightforwardly hold for an optically thin gas, in which case a~pressure-supported torus is expected to form near the equatorial Eddington capture radius. Even for an optically thick burst, the effect is important for the boundary layer \citep{Worpel2013} or the outflows in the outer region of an expanding gaseous shell \citep{Paczynski1986}. Apart from that, during a photospheric expansion burst the shell of matter may become transparent as it expands, decreasing the density. Since the equatorial drag operates on timescales much shorter than the burst duration, it may influence the late stage of an initially optically thick burst. 

Several other systems for which the under-appreciated dynamical influence of the radiation from a~compact rotating source could possibly be important are ultraluminous X-ray sources \citep{Kaaret2017}, super-giant high-mass X-ray binaries \citep{Oskinova2012} and luminous slim disks producing strong optically thin winds \citep{Dotan2011}. While the detailed consequences for realistic astrophysical contexts require more thorough considerations of the interplay between radiation, opacities, spectral energy distribution, and magnetic fields, the consequences for the analytic optically thin models are clear -- spherically symmetric models \citep[e.g.,][]{Wielgus2015b,Bollimpalli2019} do not capture, even qualitatively, the case of a~rotating luminous source and hence one should be cautious when attempting to employ them as a~framework for interpretation of observations. 

In this short paper we  focused on a~case of matter ejected from a~neutron star surface by a~near-Eddington burst of radiation. A~separate paper will be dedicated to the detailed analysis of the equation of motion components and general classification of trajectories in vicinity of a~luminous rotating neutron star (Vieira \& Wielgus, in prep).

\section*{Acknowledgements}

The author thanks Debora Lan\v{c}ov\'{a}, David Abarca, Chris Fragile, and Ronaldo Vieira for useful comments and Jae-Sok Oh for access to his derivation notes. This work was supported in part by the Black Hole Initiative at Harvard University, which is supported by a~grant from the John Templeton Foundation.




\bibliographystyle{mnras}
\bibliography{bibliography} 

\bsp	
\label{lastpage}
\end{document}